\documentclass[12pt]{article}

\usepackage{epsfig}

\def\refpar{\par\hangindent=3em\hangafter=1}
\def\reference{\relax\refpar}

\catcode`\"=\active \let"=\"
\let\3=\ss

\title{\bf SST and North American Tropical Cyclone Landfall: A Statistical Modeling Study}

\author{{Timothy M. Hall \thanks{\textit{Corresponding author address:}
Timothy Hall, NASA Goddard Institute for Space Studies, 2880 Broadway,
New York, NY, 1025. \newline{E-mail: thall@giss.nasa.gov}}}\\
NASA Goddard Institute for Space Studies, New York, NY
\and
Stephen Jewson\\
Risk Management Solutions, London, U.K.
}
\begin{document}

\maketitle

\begin{abstract}
We employ a statistical model of North Atlantic tropical cyclone (TC) tracks to investigate the relationship between sea-surface temperature (SST) and North American TC landfall rates.  The track model is conditioned on summer SST in the tropical North Atlantic being in either the 19 hottest or the 19 coldest years in the period 1950--2005.  For each conditioning many synthetic TCs are generated and landfall rates computed.  Compared to direct analysis of historical landfall, the track model reduces the sampling error by projecting information from the entire basin onto the coast.  There are 46\% more TCs in hot years than cold in the model, which is highly significant compared to random sampling and corroborates well documented trends in North Atlantic TC number in recent decades.  In the absence of other effects, this difference results in a significant increase in model landfall rates in hot years, uniform along the coast.  Hot-cold differences in the geographic distribution of genesis and in TC propagation do not significantly alter the overall landfall-rate difference in the model, and the net landfall rate is 4.7 yr$^{-1}$ in hot years and 3.1 yr$^{-1}$ in cold years.  SST influence on genesis site and propagation does modify the geographic distribution of landfall, however.  The Yucatan suffers 3 times greater landfall rate in hot years than cold, while the U.S. mid-Atlantic coast exhibits no significant change.  Landfall probabilities increase in hot years compared to all years in Florida, the U.S Gulf coast, the Mexican Gulf coast, and Yucatan with at least 95\% confidence.
\end{abstract}

\newpage
\section{Introduction}

Intense tropical cyclones (TCs) are among the most devastating of natural phenomena, and considerable effort is spent in estimating the risk of TC landfall.  Landfall risk assessments are used by the insurance industry for setting rates and by governments for establishing building regulations and planning emergency procedures.  Recently, there has been much interest in documenting and understanding trends in TC frequency and intensity, and whether such trends are related to anthropogenic climate change and increasing sea-surface temperatures (SST).  TC frequency has increased in the North Atlantic in recent years, as has the number of TCs reaching the most intense categories (Webster et al., 2005), though the global TC frequency has been approximately steady.  Theory points to a large role for SST in the maximum potential TC intensity (Emanuel, 1987), and Emanuel (2005) found a rapid increase since 1970 in global TC power dissipation that is highly correlated with SST.  Rising SST in the main development regions (MDRs) of TCs over recent decades is largely due to anthropogenic greenhouse warming (Santer et al., 2006; Elsner, 2006).

Less attention has been focused on variations in landfall risk and its geographic distribution with SST.  Changes in TC frequency are likely to affect landfall rates, but so might changes in geographic distribution of TC genesis and paths of propagation.  Lyons (2004) found that most of the difference between high and low U.S. landfall years is due to changes in the fraction of TCs making landfall, rather than changes in basin-wide TC number.  In contrast to the Atlantic-wide increase in TC power dissipation shown by Emanuel (2005), TC power dissipation at U.S. landfall has not shown any trend (Landsea, 2005), although there are many times fewer data on which to base the analysis.

The most straightforward approach to analyzing the SST dependence of landfall risk is to use historical landfall events.  This approach is sound if there are sufficient events, such as is found over large sections of coast and over many years.  However, if one aims to study changes in geographic landfall distribution and, in addition, use subsets of data years based on climate state, then sampling error becomes a major issue.   Basin-wide statistical track models are an attractive alternative.  The primary advantage of a basin-wide model is that it utilizes historical track information over the full basin, roughly 100 times more data that just at landfall.  A second advantage is that by using a track model landfall changes can be decomposed  into changes in various TC properties, such as number, genesis site, and propagation.  Here we use the track model developed by Hall and Jewson (2007a) to explore regional variations of landfall rates with SST.

In Section 2 we describe the historical data, in Section 3 we briefly review the statistical TC track model, and in Section 4 we discuss the conditioning of the model on SST.   The impact of SST on each model component is presented in Section 5.  In Section 6 we discuss the SST impact on North American landfall and relate the impact to TC number, genesis site and propagation.  We conclude in Section 7.

\section{Historical Data}

Our TC analysis is based on HURDAT data over the North Atlantic (Jarvinen et al., 1983).  We use HURDAT TCs of all intensity from 1950---2005, which encompasses 595 TCs.  Observations prior to 1950 are less reliable, as they precede the era of routine aircraft reconnaisance.  We also use SST data obtained from the UK Met Office Hadley Centre (Rayner et al., 2003).

Fig. 1 shows the evolution from 1950 to 2005 of July-August-September SST averaged over the region 290$^{\circ}$W to 345$^{\circ}$E and 10$^{\circ}$N to 20$^{\circ}$N.  There is a well-documented upward trend, particularly in the past two decades, which is largely due to anthropogenic greenhouse warming (Santer et al., 2006; Elsner, 2006).  From the 56 years we extract the hottest 19 years, with SST $>$  27.38$^{\circ}$C, and the coldest 19 years, with SST $<$ 27.07$^{\circ}$C, roughly 1/3 each of all years.   Other geographic zones for SST averaging over the tropical and mid-latitude North Atlantic result in very similar partitioning of data years.

Figs. 2a and 2b shows the historical TC tracks in the 19 cold years and the 19 hot years, of which there are 165 and 239, respectively.   In order to compute landfall rate we divide the North American continental coastline into 39 segments of varying length from Maine to the Yucatan peninsula, as shown in Fig. 3.  Mileposts A through J are shown for reference in subsequent figures.   ``Landfall'' of a TC is recorded when a track segment, heading landward, intersects a coastline segment.  A TC can make multiple landfalls.  Fig. 4a shows the total landfall counts for hot and cold years for each coastline segment, running from Maine to Yucatan.  In Fig. 4b these raw counts are converted to rates per year per 100 km of segmented coastline and plotted versus distance along the segmented coast.

The landfall counts and rates of Fig. 4 show large variations along the coast and between hot and cold.  Summed over the full coast there are 69 landfalls in 165 cold-year TCs and 110 landfalls in 239 hot-year TCs.  Some components of these hot-cold differences reflect a geophysical relationship with SST, while other components are simply due to the sampling error inherent in the low total landfall counts.  Many coastline segments have experienced only a few or no landfalls.  Our goal is to extract the components of the hot-cold differences that are geophysical by minimizing the sampling error.  In the statistical track analysis that follows we exploit the much larger TC data set over the full Atlantic basin, not just the data at landfall.  In essence, historical data from the full basin is projected onto the coastline to reduce the landfall sampling error.

\section{Basin-Wide Track Model}

Hall and Jewson (2007a) described and evaluated a statistical model of TC tracks in the North Atlantic from genesis to lysis based on HURDAT data.  The track model consists of three components: (1) genesis, (2) propagation, and (3) lysis (death).  The number of TCs in a simulation year is determined by random resampling of the historical annual TC number.  Genesis sites are simulated by sampling a kernel pdf built around historical sites.  The kernel bandwidth is optimized by jackknife out-of-sample log-likelihood maximization.  For propagation, we compute mean latitude and longitude 6-hourly displacements and their variances, by averaging of ``nearby'' historical displacements.  Nearby is defined optimally by jackknife out-of-sample log-likelihood maximization.  Standardized displacement anomalies are modeled as a lag-one autoregressive model, with latitude and longitude treated independently.  The autocorrelation coefficients are determined from ``nearby'' historical anomalies, with ``near'' optimized as above.  Finally, TCs suffer lysis with a probability determined by optimal averaging of nearby historical lysis rates.  Tracks are stochastic, and two tracks originating from the same point can follow very different trajectories.

At present no TC intensity is modeled, and no intensity information is used in the track model.  Clearly, intensity is a critical component of TC risk assessment.  In this study, however, we restrict attention to landfall by TCs of any intensity.

The advantage of a track model over direct analysis of landfall is that information is brought to bear on landfall from the full trajectories of all TCs, not just the single landfall points of landfalling TCs.  This increases the data by a factor of 100.  Not all the data is equal, and the most relevant information comes from portions of TC trajectories near the coast.  This exploitation of the most relevant data for estimating landfall is implicit in the track model.  The track-model advantage is particularly important for analysis of small coast sections; analysis of regions with low activity, where there are few or no historical landfalls; and analysis in which only subsets of data years are used, for example, conditioned on a climate state.  Hall and Jewson (2007b) compared the landfall rates of the track model to those of a Bayesian model based solely on local landfall observations, a type of statistical analysis often used in TC landfall studies.  They found that on regional and smaller scales the track model performs better in a jackknife out-of-sample evaluation with historical landfalls, despite the fact that over the entire North American coast the track model underestimates landfall by 12\%, an amount that is significant.  On a local basis, the decreased sampling error of the track model more than compensates for the increased bias.

\section{SST and Model Construction}

Or goal is to examine the effect of SST on TC landfall rates.   To do so we construct the track model separately on each historical SST set; that is, the averaging to obtain means, variances, and autocorrelation coefficients of 6-hourly track displacements, and the track lysis probabilities are restricted to the particular cold or hot subset.  The track model components are thus ``conditioned'' on data being in a particular subset based on SST.  Additionally, the kernel pdf of genesis sites is built on data only in the subset, and the annual TC number is drawn randomly only from the subset.  The hot- and cold-conditioned models are then run for a large number of years (1000), and the synthetic tracks and their landfalls analyzed.   Figs. 2c and 2d show the synthetic tracks based on the hot and cold years.  In addition, in order to isolate individual components of the SST influence on tracks we generate synthetic tracks with single model components (TC number, genesis site, propagation) conditioned on SST, while the other components are unconditioned.

In order to test the significance of hot-cold differences compared to other sources of variability we have also constructed the track model from random subsets of data, as follows:  we choose randomly a 19 year subset of the 56-year data, construct the track model, and generate 1000 synthetic tracks.  We then choose another random but non-overlapping 19-year subset, again construct the track model, and generate 1000 new synthetic tracks.  Various properties (e.g., landfall rates) of the two sets of synthetic tracks are differenced.  The selection of model components conditioned on the random 19-year periods always matches the selection of model components for SST conditioning; for example, when we condition TC propagation alone on SST we compare the hot-cold difference to random differences in which propagation alone is conditioned on the random 19-year sets, to be consistent.

The comparison of hot-cold differences to random differences calculated in this way addresses the significance of hot-cold differences compared to other sources of variability.  If the hot-cold difference of some rate stands out sufficiently from the random differences the null hypothesis, that SST has no effect on that rate, is disproved.  As we shall see, in many, but not all, regions hot-cold differences are significantly larger than random differences. Once hot-cold significance is established, however, a complementary analysis is needed to estimate the uncertainty on the estimate of a hot-year or cold-year rate.  To this end we successively remove one of the 19 years from the hot (or cold) set, for each removal conditioning the model on the 18 remaining years, generating synthetic TCs, and computing rates.  The range of resulting 18-year rates provides a measure of uncertainty of the 19-year rate.

\section{SST and Model Components}

Here we examine the impact of SST conditioning on each model component individually.  The resulting impact on landfall rates, both individually and in combination, is presented in Section 6.

\subsection{TC Number}

Our model for annual TC number is simple, consisting of random resampling from the TC annual numbers in the historical years of the data-year subset in question (e.g., hot, cold, or random).  The mean annual TC number over 1000 simulated hot years is 12.8 yr$^{-1}$, while for cold years it is 8.5 yr$^{-1}$.  The hot-cold difference, 4.3 yr$^{-1}$ (50\%), is highly significant.  By comparison, the mean absolute-magnitude of differences across the 20 random 19-year pairs is 0.8 yr$^{-1}$, with an rms deviation of 0.6 yr$^{-1}$.  In fact, none of the 40 random 19-year subsets has as few TCs as the cold years and none has as many TCs as the hot years.  As we shall see below this difference in TC number has the single greatest influence on landfall rates.  By construction the effect of number on TC tracks and landfall has no geographic structure.

\subsection{Genesis Site}

Given an annual number of TCs, where should they originate?  Figs. 5a and 5b show the kernel pdfs of genesis sites for the cold years and the hot years.  Not surprisingly, the hot years have greater probability for genesis in the MDR.  Ocean heat is a key ingredient to cyclogenesis and TC intensification (Emanuel, 1987), and the MDR is precisely the region whose SST we have used to condition the model genesis.  This highlights the fact that our separation of TC number and genesis site is to some degree artificial: much of the hot-year number increase is concentrated in the defining region of hot years.  However, there is additional hot-cold genesis-site structure outside the MDR.  Hot years also have greater probability for genesis off the Mexican Gulf Coast and just east of the Yucatan peninsula.   In contrast, individual events in cold years have greater probability for genesis in the northern Gulf of Mexico and in the western Atlantic off the southeastern U.S coast.  For comparison, Fig. 5c shows the genesis pdf for all 56 data years, which is intermediate.  (We emphasize that these pdfs are normalized spatially.  They provide the probability of a genesis site, given a genesis occurrence.  If a pdf increases in one region, it must decrease elsewhere.)

Fig. 5d shows shows a measure of the significance of the hot-cold difference in genesis-site pdfs.  The pdfs of the 20 random 19-year pairs are differenced and the rms deviation of the  differences computed at each plotted point in the basin ($1^{\circ} \times 1^{\circ}$).  The hot-cold difference is plotted where (1) its magnitude is greater than one standard deviation across the random differences (i.e., it is significant), and (2) its magnitude is within 50\% of the maximum hot-cold magnitude (i.e., it is large).  The differences noted qualitatively comparing Fig. 5a to 5b are confirmed to be significant.

The hot-cold differences in genesis site impact the distribution of modeled TC tracks throughout the basin and also affect the modeled TC landfall rate.  Landfall rates are analyzed in Section 6.  Here, we determine the impact of genesis site changes on TC tracks by analyzing the number of TCs crossing lines of constant longitude and latitude.  For this purpose, only genesis site pdfs are conditioned on SST, while TC number and propagation are unconditional.  Fig. 6 shows the zonal ``flux'' of hot- and cold-year TCs; that is, the number of TCs crossing lines of constant latitude per year counted in 5$^{\circ}$ latitude bins.  Eastward and westward fluxes are shown separately, as are the eastward and westward hot-cold flux differences.  Plotted with the flux differences are the random-difference standard deviations as a measure of statistical significance.  Fig. 7 shows a comparable plot for the meridional TC flux.

The greater probability of genesis in the MDR in hot years results in more TCs propagating westward in the subtropics.  The hot-cold difference is significant.  Many TCs then curve northward, followed by eastward.  A significantly greater hot-year northward flux is seen across 20$^{\circ}$N.  The eastward midlatitude TC flux is greater in hot years, but the hot-cold difference is in most places not significant compared to the random differences.   By contrast, the influence of the cold-year genesis-site peak off the southeastern U.S. coast can be seen in the significantly greater cold-year northward flux off the U.S. coast through 30$^{\circ}$N and 40$^{\circ}$N.

\subsection{TC Propagation}

We now analyze the impact of SST on TC propagation by conditioning on SST the mean 6-hourly displacements, the variance about the mean, the autocorrelation, and the lysis.  The genesis-site pdfs and the TC numbers are unconditional.  First, to illustrate the SST influence we generate 1000 hot and cold synthetic TCs from the same genesis site and compute the density of track points.  Fig. 8a and 8b shows the resulting regions of high track-point density from eastern and western genesis sites in the MDR.  The hot-cold TC difference is most apparent for the western genesis site.  Hot-year TCs are more likely to bend northward and eastward earlier, while cold-year TCs tend to propagate further west, including into the Caribbean, before heading north.

The zonal and meridional fluxes are shown in Figs. 9 and 10.  These figures are analogous to Figs. 6 and 7, but now for the case of SST-conditioned propagation and lysis, but unconditional TC number and genesis site.  Consistent with Fig. 8 westward cold-year flux is greater than hot-year flux through 290$^{\circ}$E, although the difference is only marginally significant.  By contrast the northward hot-year flux is greater through 20$^{\circ}$N, 30$^{\circ}$N, and 40$^{\circ}$N in the western Atlantic.  The most significant propagation differences are seen  in the eastward flux in midlatitudes, which is significantly greater in hot years than cold.

The physical mechanisms relating MDR SST to TC propagation are unclear.  It may be that the signal seen here is a projection of the North-Atlantic Oscillation (NAO) on our SST data-year partitioning.  In the NAO's high phase SST in the tropical North Atlantic is lower, and the subtropical surface-pressure high extends further west (Wang, 2002).  This would tend to cause TCs to propagate further west before curving north.  During the NAO's low phase the SST is higher, and is therefore associated with more TCs (Elsner et al., 2006), but the subtropical high does not extend as far west, and TCs curve north sooner.

\section{SST and Landfall Rates}

We now analyze the landfall-rate characteristics of the synthetic TC data.   Figs 11--14 illustrate landfall rates (counts per year per 100 km) and rate differences as a function of distance along the segmented coastline running southward from Maine, as in Fig. 4. The general features of the model's landfall rate and its evaluation against historical rates are discussed in Hall and Jewson (2007a; 2007b).  Accumulated over the full coast the model suffers from a low landfall bias of about 12\% compared to the 1950--2005 historical rate, a difference which is marginally significant compared to the rms deviation across many 56-year simulations.

Landfall rates show strong geographic variations, with several local maxima.  The rates are sensitive to the orientation of coast segments, with segments perpendicular to the local mean TC path experiencing more landfalls.  Hence, there is a local landfall-rate minimum off the southeast U.S. Atlantic coast (between mileposts C and D), where the coast segments are close to parallel with the mean TC tracks, while there is a maximum near Cape Hatteras (mileposts B--C) and Long Island (mileposts A--B), where south-facing coast segments jut out into oncoming TCs.  One implication is that different segmented models of the coastline will result in different distributions, making it difficult to compare directly the quantitative landfall rates from one study to another.

\subsection{SST-Conditioned TC Number}

Fig. 11 shows the landfall rates and the hot-cold difference for the case with only TC number conditioned on SST.  TC-number conditioning on SST has no geographic structure, and the hot-cold landfall-rate ratio is roughly uniform, with the hot-year rate everywhere significantly greater than the cold-year rate.  (The ratio is uniform in the limit of large number of simulation years.)  Accumulated over the entire segmented coast there are 3.2 cold landfalls per year and 4.8 hot landfalls per year, a ratio that simply reflects the historical  hot-cold TC-number ratio 239/165.  The hot-cold landfall-rate difference, 1.5 yr$^{-1}$, is highly significant.  Across the 20 random 19-year pairs the rms deviation of landfall-rate differences is only 0.4 yr$^{-1}$.

\subsection{SST-Conditioned Genesis Site}

Fig. 12 is comparable to Fig. 11, but now for the case of genesis-site pdf conditioned on SST, while the other components are unconditional.  Accumulated over the entire coastline the hot-year landfall rate is 3.9 yr$^{-1}$ compared to the cold-year rate 3.8 yr$^{-1}$, a difference that is well within the range of random differences, which has an rms deviation of 0.5 yr$^{-1}$.  A few regions, however, have significant hot-cold differences.  On the Mexican Gulf coast (milepost G) and the eastern Yucatan (milepost J), hot-year genesis sites result in significantly greater landfall, consistent with the hot-year genesis-site peaks just off the coasts of these regions (Fig. 5).  By contrast, on the northern U.S. Gulf coast (between mileposts E and F) the cold-year landfall rate is significantly higher, consistent with the cold-year genesis-site peak in the central and northern Gulf of Mexico.  In addition, the peak in cold-year genesis sites off the southeastern U.S. coast causes cold-year landfall in the mid-Atlantic U.S. coast to be greater (between mileposts B and C), a difference that is marginally significant.

Note that these landfall rate differences do not take TC intensity into account.  Cold-year genesis sites cause more cold-year landfalls on the U.S. mid-Atlantic coast, but these landfalls are not far from their genesis sites, and have not had much time to intensify.  By contrast, the more frequent hot-year genesis in the MDR results in lower probability per TC of making landfall---there is more opportunity to veer away from coast---but also allows more time for TCs to intensify.

\subsection{SST-Conditioned Propagation and Lysis}

Fig. 13 shows the landfall rates for SST conditioning of propagation and lysis, with the genesis-site and TC number unconditional.  Accumulated over the entire coastline the hot-year landfall rate is 3.7 yr$^{-1}$ and the cold-year landfall rate is 3.9 yr$^{-1}$, a difference that is well within the random variability of random differences, which now has an rms deviation of 0.6 yr$^{-1}$.  Once again, however, on some regions the hot-cold differences are significant, albeit marginally so.  On the U.S. northeast and mid-Atlantic coast segments facing the oncoming mean TC tracks (Long Island and Cape Hatteras) cold-year landfall rates are greater, reflecting the fact that in cold years TCs propagate further west before veering north (Figs 8--10).  Note in Fig. 10 that across 40$^{\circ}$N just off the U.S. coast the northward flux of TCs is greater in cold years than hot (marginally significant), even though only a few degrees further east the opposite is true.  By contrast on the eastern Yucatan (milepost J) the SST effect on propagation causes hot-years to have greater landfall rate.  This effect must be well localized, because there is little hot-cold difference in the westward TC flux across 280$^{\circ}$E, about 10$^{\circ}$ east of Yucatan.

We have not separated the SST influence on propagation and lysis, in order to keep the number of figures manageable.  Separate analysis (not shown) indicates no significant hot-cold landfall-rate differences due to lysis alone.

\subsection{Full SST Conditioning}

Fig. 14 shows the hot-year and cold-year landfall rates when all model components are conditioned on SST simultaneously.  Accumulated over the entire coastline the SST effects on genesis site and propagation are negligible, and the fully-conditioned hot-cold landfall rates are nearly identical to the TC-number only case: 3.1 yr$^{-1}$ in cold years and 4.7 yr$^{-1}$ in hot years.  The difference, 1.5 yr$^{-1}$, is significant, about 2 times the random difference rms deviation of 0.7 yr$^{-1}$.   Note that compared to the TC-number conditioning, the random variation is now larger, reflecting variability in more model components.

Hot-cold differences in genesis site and propagation affect regional landfall rates, however, in some places amplifying the TC-number effect and in other places ameliorating it.  On the eastern Yucatan coast the landfall effects of SST on TC number, genesis site and propagation all have the same sign: greater hot-year than cold-year landfall.   With all components conditioned on SST the Yucatan landfall rate is about 3 times higher in hot years than cold, a difference which is highly significant.  By contrast, the greater landfall rates in hot years on the northeast and mid-Atlantic U.S. coast due to TC number alone is countered by the greater cold-year rates due to genesis site and propagation.  The net effect is no significant hot-cold landfall-rate difference in this region.  Most of the SST effect via TC-number on Florida and the U.S. Gulf Coast survives under full conditioning, although its significance is now marginal, as the random variability is larger.   Note that significance increases when landfall is accumulated over larger regions.  Table 1 lists the hot-year and cold-year rates on the 6 regions indicated in Fig. 3.  Also listed are the $Z$ scores of the differences; that is, the hot-year minus cold-year rate difference divided by the rms deviation of the random differences.   On the U.S. northeast and mid-Atlantic coastlines the hot-cold differences are not significant.  On Florida, the U.S Gulf coast and the Mexican Gulf coast the hot-cold differences are marginally significant ($Z$ scores of 0.9, 1.1, and 0.9, respectively).  On the Yucatan peninsula the hot-cold difference is highly significant, with $Z=2.4$.

\subsection{Landfall Fraction Variability on U.S. Coast}

The fact that genesis-site and propagation conditioning on SST has little effect on landfall rates accumulated over large sections of the coast is reflected in the small hot-cold difference in landfall fraction (fraction of TCs making landfall).  Over the the U.S. portion of the coastline (segments 1--23) the modeled landfall fractions are 0.28 in cold years and 0.29 in hot years.  The actual historical record reveals a larger difference: 0.30 in cold years and 0.35 in hot years (50 U.S. landfalls in 165 historical cold-year TCs and 84 U.S. landfalls in 239 historical hot-year TCs).  Here we examine this apparent discrepancy.

The historical hot and cold years subsets analyzed here each have 19 years.  We divide the 1000-year hot and cold simulations into 52 19-year periods and compute the landfall fractions in each period.  These fractions exhibit considerable variability.  Fig. 15a shows the pdf of these landfall fraction for hot and cold years, along with the total simulated hot and cold fractions and the historical hot and cold fractions.  The historical fractions each fall within the variability of the simulated fractions.  We then take many random hot-cold differences of landfall fractions and compute the pdf of the differences (Fig. 15b).  The historical difference (hot $>$ cold by 0.05) is well within the range of random differences in the simulations.  This suggests that the apparent discrepancy between the historical and simulated hot-cold landfall fraction differences is not significant.  It also suggests that no significant impact of SST on U.S. landfall fraction can be gleaned from observations over the period 1950--2005.

\subsection{Landfall Probabilities}

We have demonstrated that for many features and on many regions the landfall rates in hot years are significantly greater than those in cold.  We now convert the hot-year rates to landfall probabilities with confidence limits and compare to the same probabilities in all years.  Landfall can be modeled as a Poisson process (Bove et al., 1998) whose rate is uncertain, due to the finite historical record.  (We have verified that the distributions of annual landfall counts in the model are Poisson.)  Incorporating this uncertainty into the landfall probability results in a negative binomial distribution (Elsner and Bossak, 2001; Hall and Jewson, 2007b).  Given $i$ landfalls in $m$ years the probability of $n$ landfalls in a subsequent year is

\begin{equation}\label{e:prob}
f(n|i) = \frac{(i+n)!}{i!n!} \left( \frac{m}{m+1} \right)^{i+1} \left( \frac{1}{m+1} \right).
\end{equation}
Expression (\ref{e:prob}) reduces to a Poisson distribution for large $m$ and $i$.  In the case of the synthetic TCs  $m = 1000$, and $i$ on 100 km coastline sections varies from 1 to 144 (using all years), depending on region.

Fig. 16a shows the probability of at having at least one landfall in a year in 100 km sections along the segmented coastline for the hot years and for all years, based on these distributional assumptions.  Fig 16b shows the comparable distribution for the probability of at least two landfalls.  Plotted for the hot-years are the 90\% confidence limits of the probability determined from a distribution of probabilities.  The distribution is determined, in turn, by successively dropping out one of the 19 hot years, conditioning the model on the remaining 18 hot years, generating synthetic TCs, and computing landfall rates and probabilities.  For example, on segment 19 (295 km in length facing southeast of the Louisiana coast between mileposts E and F in Fig. 3) there is a probability of 0.085--0.117 of at last one landfall in hot years in 100 km sections, versus a probability of 0.065 in all years, an increase of 31\% to 80\%.  For two or more landfalls there is a probability of 0.0038--0.0072 for hot years versus 0.0022 for all years, an increase of  73\% to 327\%.  By contrast, on segment 8 (91 km in length facing south from Cape Hatteras between mileposts B and C) the probability range for at least one landfall in 100 km is 0.110--0.152, which encompasses the probability for all years of 0.134.  The probability here for at least two landfalls in hot years is 0.0064--0.0122, which again encompasses the probability, 0.0094, for at least two landfalls in all years.

Listed in Table 2 are the hot-year and all-year landfall probabilities on 6 larger regions of coast: the U.S. northeast, the U.S. mid-Atlantic, the Florida peninsula, the U.S. Gulf coast (excepting Florida), the Mexican Gulf coast (excepting Yucatan), and the Yucatan peninsula.  Mileposts delineating these regions are indicated in Fig. 3.  Compared to Fig. 16, here the probabilities are accumulated over the entire region, rather than in 100 km sections.  On the U.S. northeast coast there is apparently a slight reduction in landfall probability in hot years compared to all years.  However, as indicated in Table 1, the hot-cold difference is not significant, and we conclude that there is no change in landfall probability in hot years.  On the U.S. mid-Atlantic coast the hot-year and all-year probabilities are not different, and the hot-cold difference is within the range of random differences.  On Florida, the U.S. Gulf Coast, and the Mexican Gulf Coast the hot-cold differences are marginally significant ($Z = 0.9--1.1$), while on the Yucatan coast the hot-cold difference is highly significant ($Z=2.4$).  On Florida, the U.S. Gulf, and the Yucatan peninsula, the hot-year landfall probabilities are higher than the probabilities for all years, with at least 95\% confidence (i.e., the lower 90\% confidence limit of hot years is greater than the probability for all years). d  The increase in landfall probability in hot years is greatest on the Yucatan peninsula; e.g., there is a 90\% to 250\% increase in probability of two or more TCs making landfall in hot years compared to all years.  On the U.S. Gulf coast the comparable increase is 14\% to 36\%.  The hot-year increase on the U.S. Gulf coast for making three or more landfalls is 20\% to 70\%.

\subsection{Sensitivity to SST Averaging Region}

We have performed additional limited analysis with alternative spatial and temporal averaging of SST to test the robustness of our conclusions.  Increasing the latitude range from 10$^{\circ}$--20$^{\circ}$N to as much as 0$^{\circ}$--40$^{\circ}$N, the longitude range from 290$^{\circ}$--345$^{\circ}$E to as much as 270$^{\circ}$--345$^{\circ}$E, and altering the time period from Jul--Sep to Aug--Oct in various combinations never changes the 19 hottest and 19 coldest years among 1950--2005 by more than 3 years.  Thus, our TC model conditioning will not be strongly affected.  For the case 10$^{\circ}$--30$^{\circ}$N, 270$^{\circ}$--345$^{\circ}$E, and Aug--Oct (change in 3 of the 19 hottest and coldest years) we have fully conditioned the model and simulated 1000 hot- and cold-year TC tracks.  The genesis-site pdfs are altered quantitatively, though not qualitatively, and the hot-cold landfall-rate difference over the entire coastline is reduced (4.5 yr$^{-1}$ in hot years, 3.5 yr$^{-1}$ in cold years).  However, the basic features described above are unchanged.  There is no significant hot-cold difference for the U.S. NE and mid-Atlantic coasts, while the remaining regions exhibit significant hot-cold differences.

\section{Conclusions}

We have used a statistical tropical cyclone (TC) track model to elucidate the relationship between tropical Atlantic SST and TC landfall rates on North America.  The advantage of a track model over direct analysis of historical landfall rates is the reduction in sampling error from using the much larger quantity of data over the entire ocean basin.  This reduced sampling error allows meaningful examination of detailed geographic structure in landfall rate and its dependence on SST.  We have constructed the model components (annual TC number, genesis site, and propagation) individually and together from the 19 lowest and 19 highest SST years in 1950--2005, generated 1000s of synthetic TC tracks, and analyzed their landfall characteristics.  We have constructed the model on randomly selected pairs of 19-year periods and compared the range of landfall differences among these pairs to the hot-cold differences as a test of hot-cold significance.  In addition, we have constructed the model on sub-samples of the 19 hot years in order to estimate confidence limits on the hot-year rates.

TC number has a large and well known dependency on SST: there are 45\% more TCs in the hot years than cold, a difference which is highly significant compared to random sampling.  Alone, this SST dependency of TC number translates to an identical fractional landfall-rate increase, uniform along the coast. Possibly modifying this TC-number effect is variation with SST of the geographic distribution of genesis and the propagation of TCs.  It turns out, however, that accumulated over the entire coastline there is little additional effect on landfall of genesis-site and propagation variations.  When all components of the model are conditioned on being in either the cold or hot SST years we find a landfall rate of 4.7 yr$^{-1}$ in hot years and 3.1 yr$^{-1}$ in cold years, rates whose ratio is not significantly different than the historical hot-cold TC-number ratio 239/165.  Over the period 1950--2005 TCs in the 19 hot years have a higher U.S. landfall fraction (0.35) than TCs in the 19 cold years (0.30), but we have argued that this difference is not significant.

There are, however, significant differences in the geographic distribution of landfall in hot and cold years.  TC number alone causes a uniform increase in landfall rate in hot years.  Changes in genesis site and propagation amplify this increase in some regions, most notably the eastern Yucatan peninsula, whose landfall rate is 3 times higher in hot years than cold.  In contrast, genesis-site and propagation dependence on SST ameliorate the increase in landfall on the U.S. northeast and mid-Atlantic coast, with a net result of no significant hot-cold difference in landfall rate.  Landfall rates on the Florida and northern U.S. Gulf Coasts are higher in hot years than cold, although the difference is less significant than would be concluded from TC number alone.  We have also computed the hot-year increase in landfall probabilities compared to all years with confidence limits determined from subsampling the 19 hot data years.  Landfall probabilities on Florida, the northern U.S. Gulf coast, and the Yucatan peninsula are higher in hot years with at least 95\% confidence.  We have verified that these qualitative features of regional hot-cold differences are not sensitive to the SST averaging.

It is important to note that no intensity model is included here, nor is any intensity used in the analysis.  Our estimated landfall rates refer to any named TC.  Overall, landfall rates are, of course, lower if attention is restricted to the most intense TCs.  In addition, regional hot-cold differences might be quite different.  For example, cold-year genesis sites cause greater cold-year landfall rates on U.S. mid-Atlantic coast.  But these landfalls are not far from their genesis sites, and so have not had much time to intensify.  By contrast, more frequent genesis in the tropical eastern Atlantic in hot years may result in a lower chance per TC of making landfall---there is more opportunity to veer away from coast---but the TCs that do make landfall have had more time to intensify.

Increases in SST in regions of tropical cyclogenesis over recent decades is due primarily to anthropogenic greenhouse forcing (Santer et al., 2006; Elsner, 2006).  These increases are likely to persist and to intensify because (1) the ocean has large thermal intertia, so that (2) the ocean is still responding to greenhouse forcing of the last few decades (Hansen et al., 2005), and (3) greenhouse gas forcing will almost certainly continue to grow over the next few decades.  It must be acknowledged that the mechanisms for cyclogenesis and TC intensification are poorly understood, and factors in addition to SST play important roles.  TC frequency in other ocean basins has not increased as it has in the Atlantic.  Moreover, the association we see between SST and TC propagation is not understood, and may not be directly causal; i.e., there may be an unidentified mechanism responsible for variability in both SST and propagation.  However, our results show a significant increase of North American TC landfall with tropical SST in rough proportion to the increase in TC number.  The implication is clear for higher TC landfall risk with future SST increases.

Finally, we note that the vast majority of the upward trend in financial loss due to TC landfall is explained by increased exposure; that is, increased coastal population and development (Pielke and Landsea, 1998).  It is sometimes argued that this fact trivializes any climate-change impact on TC frequency and intensity.  The argument is wrong, in our view.  Regardless of past damages, coastal regions will continue to be populated and developed.  To plan development, to plan emergency procedures, and to set insurance rates developers, governments and insurers need to know the risk of TC damage and its potential for future change.   This is the question that is addressed by this study and other studies seeking to understand the connection between TCs and climate variation, including secular anthropogenic change.

\section{Acknowledgments}
NASA is acknowledged for support of this research.  We also thank NOAA's Atlantic Oceanographic and Meteorological Laboratory for maintenance of the HURDAT data base.
\clearpage

\section{References}

\noindent

\reference
Bove, M. C., J. B. Elsner, C. W. Landsea, and J. J. O'Brien, 1998, Effect of El Nino on U.S. landfalling hurricanes, revisited, {\it Bull. Amer. Meteor. Soc.}, {\bf 79}, 2477--2482.

\reference
Elsner, J. B., 2006, Evidence in support of the climate change-Atlantic hurricane hypothesis, {\it Geophys. Res. Lett.,} {\bf 33}, doi:1029/2006GL026869.

\reference
Elsner, J. B., 2006, Forecasting U.S. hurricanes 6 months in advance, {\it Geophys. Res. Lett.}, {\bf 33}, doi:1029/2006GL025693.

\reference
Elsner, J. B., and B. H. Bossak, 2001, Bayesian analysis of U.S. hurricane climate, {\it J. Clim.}, {\bf 14}, 4341--4350.

\reference
Emanuel, K., 1987, The dependence of hurricane intensity on climate, {\it Nature}, {\bf 326}, 483--485.

\reference
Emanuel, K., 2005, Increasing destructiveness of tropical cyclones over the past 30 years, {\it Nature,} {\bf 436}, 686--688.

\reference
Hansen, J. E., et. al., 2005, Earth's energy imbalance: Confirmation and implications, {\it Science,} {\bf 308}, 1431--1435.

\reference
Hall, T. M., and S. Jewson, 2007a: Statistical modeling of North Atlantic tropical cyclone tracks, {\it Tellus,} in press.

\reference
Hall, T. M., and S. Jewson, 2007b: Comparison of local and basin-wide methods for risk assessment of tropical cyclone landfall, {\it J. Appl. Meteorol. Clim.}, submitted.

\reference
Jarvinen, B. R., C. J. Neumann, and M. A. S. Davis, 1984: A tropical cyclone data tape for the North Atlantic Basin, 1886--1983, contents, limitations, and uses, {\it NOAA Tech. Memo.}, NWS NHC 22, Miami, Florida.

\reference
Landsea, C., 2005, Brief communications: hurricanes and global warming, {\it Nature}, {\bf 438}, E11--E12.

\reference
Lyons, S. W., 2004, U.S. tropical cyclone landfall variability: 1950--2002, {\it Wea. Forecasting}, {\bf 19}, 473--480.

\reference
Pielke, Jr., R. A., and C. W. Landsea, 1998, Normalized hurricane damages in the United States, {\it Wea. Forecasting}, {\bf 13}, 621--631.

\reference
Rayner, N. A., et al., 2003, Global analysis of sea-surface temperature, sea ice, and night marine air temperature since the late nineteenth century, {\it J. Geophys. Res.}, {\bf 108}, doi:10.1029/2002JD002670.

\reference
Santer, B. S., et al., 2006, Forced and unforced ocean temperature changes in Atlantic and Pacific tropical cyclogenesis regions, {\it Proc. Nat. Acad. Sci.}, {\bf 103}, 13905--13910.

\reference
Wang, C., 2002, Atlantic climate variability and its associated atmospheric circulation cells, {\it J. Clim.}, {\bf 15}, 1516--1536.

\reference
Webster, P. J., G. J. Holland, J. A. Curry, and H.-R. Chang, 2005: Changes in tropical cyclone number, duration, and intensity in a warming environment.  {\it Science}, {\bf 309}, 1844--1846.

\clearpage

% Figure 1:
\begin{figure}
\noindent\includegraphics[scale=0.85]{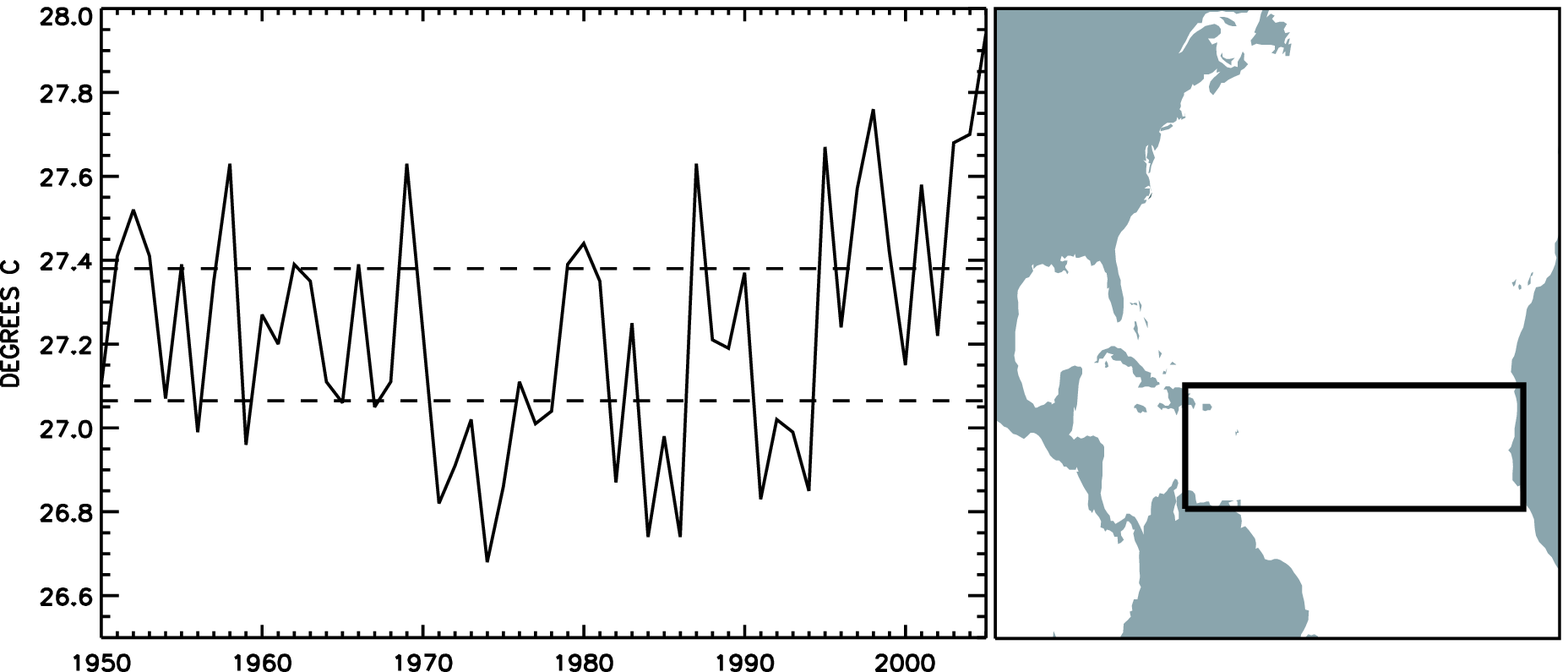}\\
\caption{History of July-August-September SST (left) averaged over the region 345$^{\circ}$E to 290$^{\circ}$E and 10$^{\circ}$N to 20$^{\circ}$N from the Hadley Centre SST data (Rayner et al., 2003).  The horizontal dashed lines indicate the thresholds for hot and cold years (27.38$^{\circ}$C and 27.07$^{\circ}$C) used for model conditioning. The averaging region is plotted in the map (right).}
\label{f1}
\end{figure}

% Figure 2:
\begin{figure}
\noindent\includegraphics[scale=0.8]{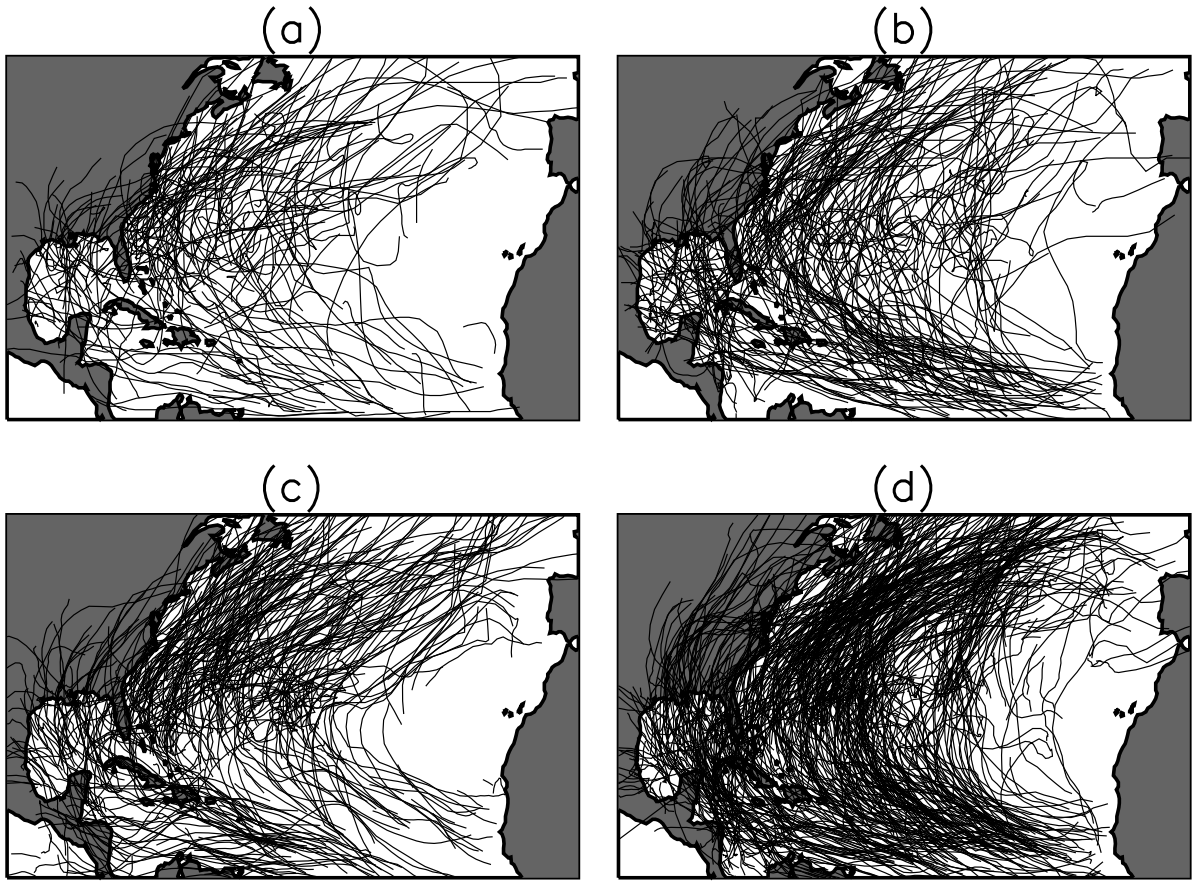}\\
\caption{(a) Historical TC tracks in cold years and (b) hot years. There are 165 cold-year TCs and 239 hot-year TCs in the 56-year period 1950--2005. (c) One thousand years of simulated TCs conditioned on cold years and (d) hot years.  There are 8725 cold-year simulated TCs and 12739 hot-year simulated TCs.  Only every 25th simulated TC is shown for clarity.  Subsequent figures address detailed differences between the cold-year and hot-year simulations.}
\label{f1}
\end{figure}

% Figure 3:
\begin{figure}
\noindent\includegraphics[scale=1.3]{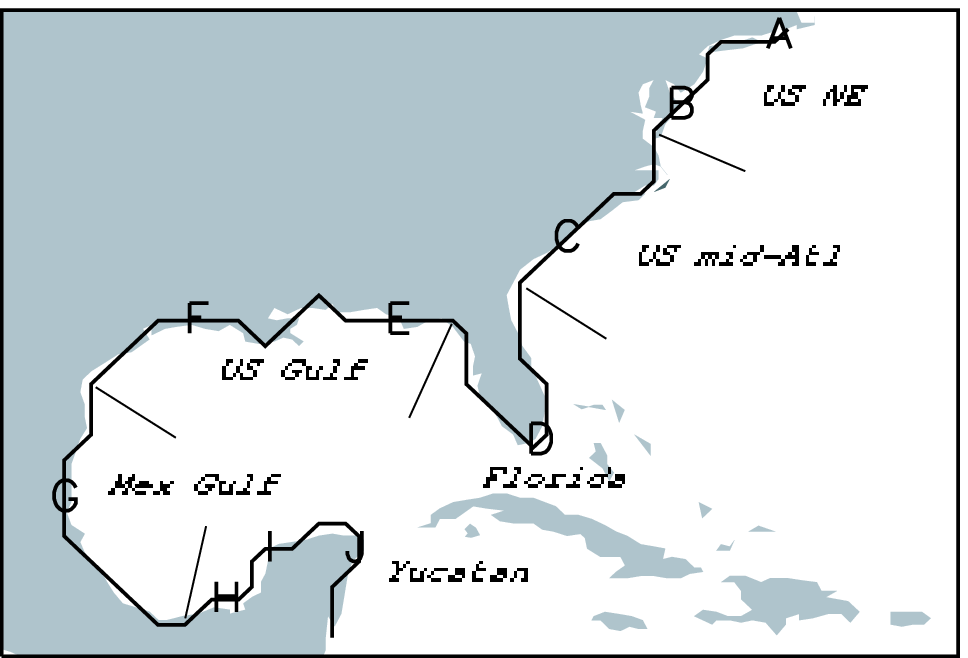}\\
\caption{Eastern North American coastline divided into 39 segments.  Select segments are labeled with letters for reference in subsequent figures. Line segments emanating from the coast delineate larger coastal regions, as labeled.}
\label{f1}
\end{figure}

% Figure 4:
\begin{figure}
\noindent\includegraphics[scale=1]{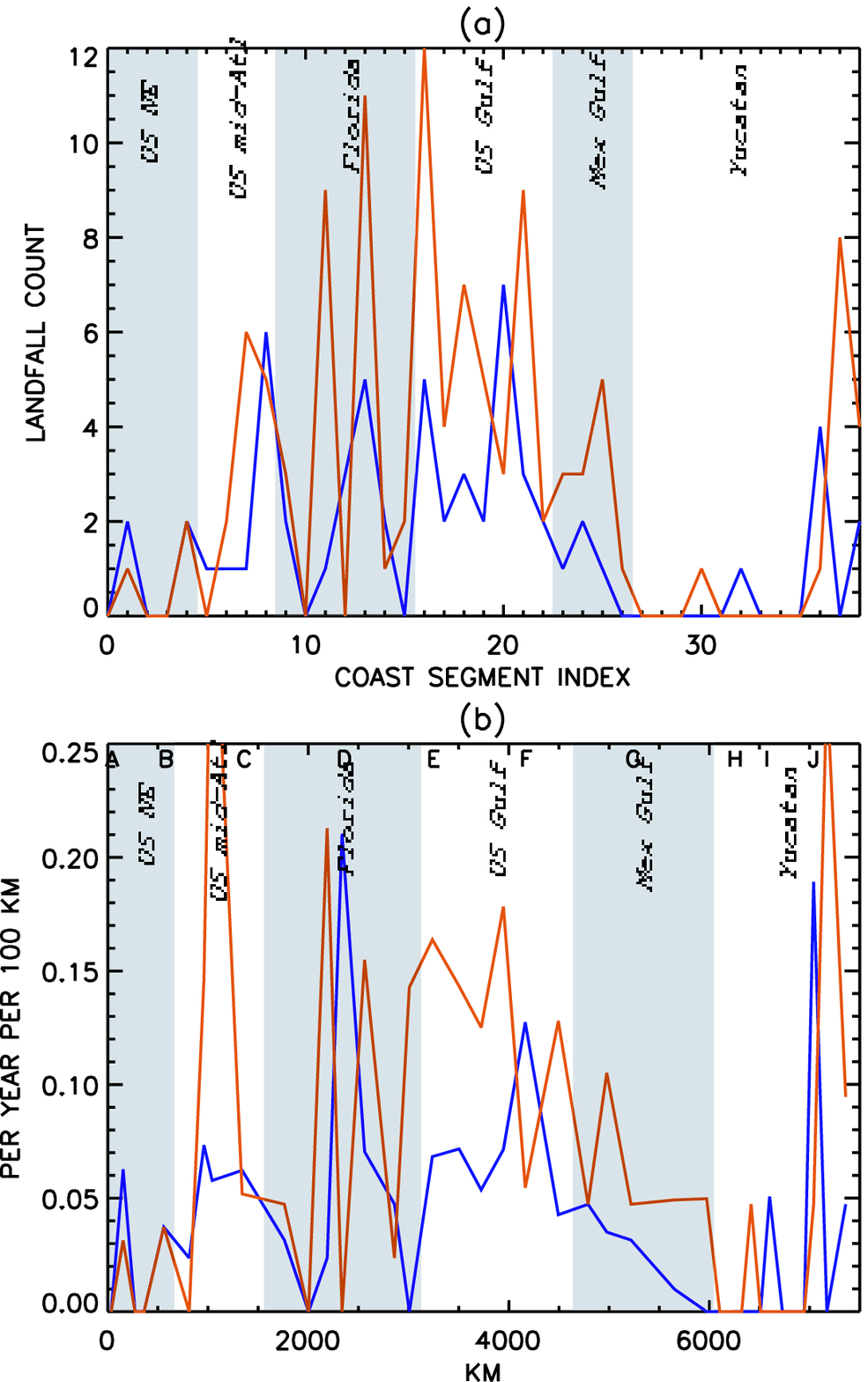}\\
\caption{(a) Total historical TC landfall counts on each coastline segment for the 19 hot years (red) and 19 cold years (blue).  (b) The landfall rates (landfalls per year per 100 km of segmented coastline) derived from the counts.  Mileposts (A--J) and regions are labeled for ease of reference to Fig. 3.  Vertical shaded regions correspond to the regions defined in Fig. 3.}
\label{f1}
\end{figure}

% Figure 5:
\begin{figure}
\noindent\includegraphics[width=38pc,angle=0]{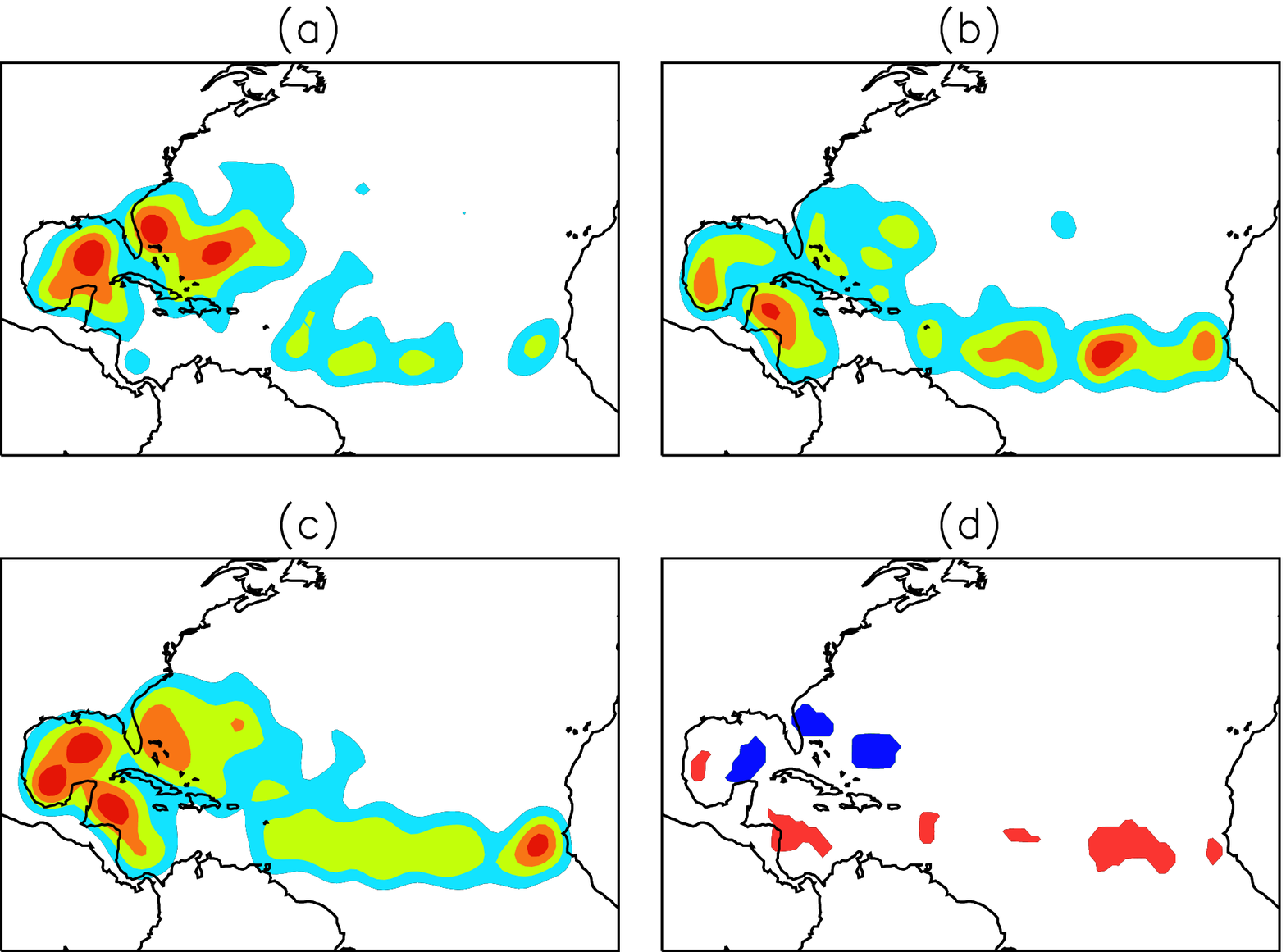}\\
\caption{Genesis site pdfs for (a) cold years and (b) hot years.  The pdfs are scaled to unit maxima and the contour intervals are 0.2, 0.4, 0.6, and 0.8.  (c) The unconditioned genesis-site pdf.  (d) The hot-cold differences.  Values are only plotted in (d) where the hot-cold difference is both large and significant: Red indicates regions where hot $>$ cold by an amount at least 50\% of the maximum difference and the difference is greater than one rms deviation across the random differences.  Blue indicates regions with the same magnitude and significance thresholds, but for cold $>$ hot.}
\label{f1}
\end{figure}

% Figure 6:
\begin{figure}
\noindent\includegraphics[width=38pc,angle=0]{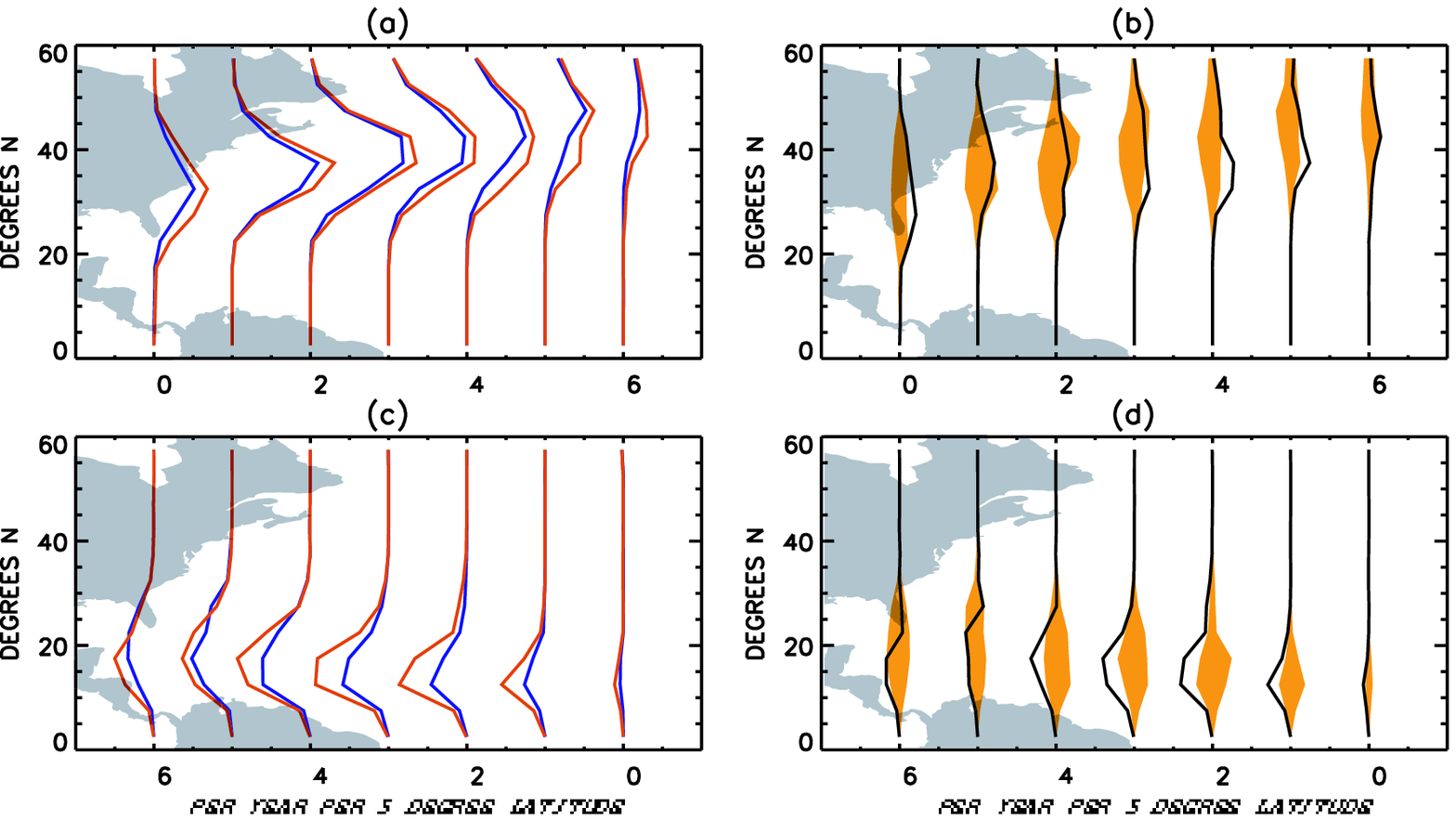}\\
\caption{TC ``flux'' across 7 lines of constant longitude (280$^{\circ}$W through 340$^{\circ}$W every 10$^{\circ}$ from left to right) as a function of latitude (vertical axes).  In this figure SST conditioning is performed for the genesis-site pdfs only; i.e., there is no conditioning for the TC number, propagation or lysis. Units are numbers per year of TCs crossing the longitude lines accumulated in 5$^{\circ}$ latitude bins. Curves are shifted 10$^{\circ}$ for each longitude line. Shown are (a) the eastward TC fluxes for cold-years (blue) and hot-years (red); (b) the difference (solid lines) of the hot-cold eastward fluxes; (c) the westward fluxes; and (d) the difference of the hot-cold westward fluxes.  To test significance of the hot-cold differences, also shown (orange filled) in (b) and (d) are the mean $\pm$ rms deviation of the eastward and westward random flux differences.}
\label{f1}
\end{figure}

% Figure 7:
\begin{figure}
\noindent\includegraphics[scale=1]{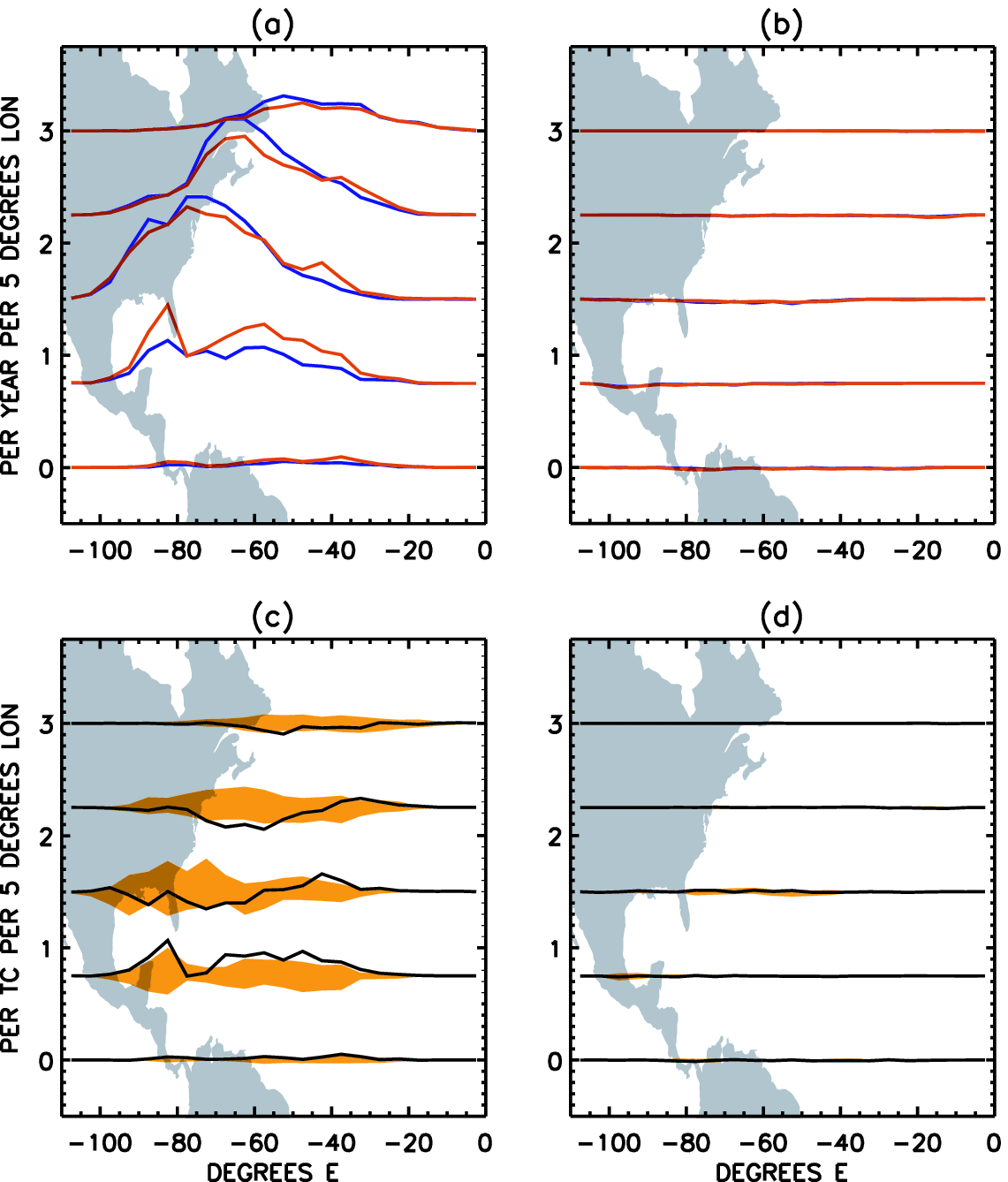}\\
\caption{TC ``flux' across 5 lines of constant latitude (10$^{\circ}$N to 50$^{\circ}$N every 10$^{\circ}$ latitude, bottom to top) as a function of longitude (horizontal axes).  In this figure SST conditioning is performed for the genesis-site pdfs only; i.e., there is no conditioning for the TC annual number, propagation or lysis. Units are numbers per year of TCs crossing the latitude lines accumulated in 5$^{\circ}$ longitude bins. Curves are shifted 10$^{\circ}$ for each latitude line. Shown are (a) the northward TC fluxes for cold-years (blue) and hot-years (red); (b) the southward fluxes; (c) the difference (solid lines) of the hot-cold northward fluxes;  (d) the difference of the hot-cold southward fluxes.  To test significance of the hot-cold differences, also shown (orange filled) in (c) and (d) are the mean $\pm$ rms deviation of the northward and southward random flux differences.}
\label{f1}
\end{figure}

% Figure 8:
\begin{figure}
\noindent\includegraphics[width=38pc,angle=0]{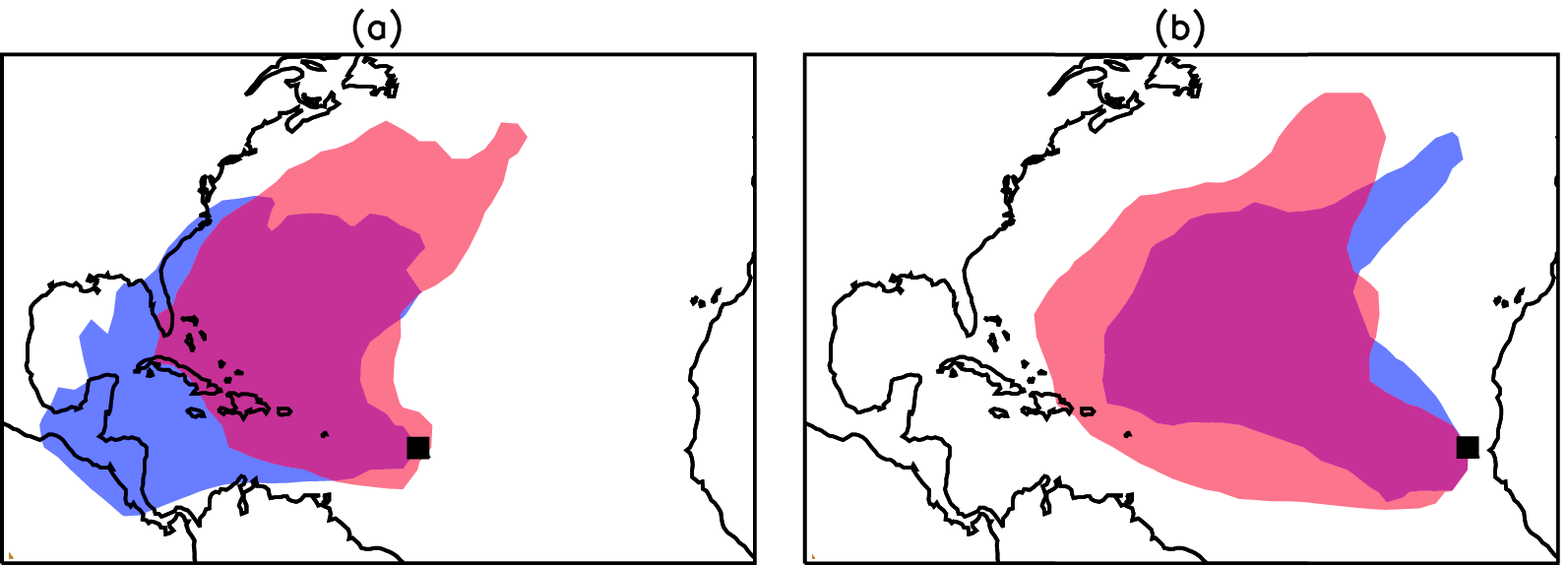}\\
\caption{Regions of high track-point density for hot years (red) and cold years (blue) for TCs originating at (a) 310$^{\circ}$E, 15$^{\circ}$N and (b) 340$^{\circ}$E, 15$^{\circ}$N.  Here, only the propagation and lysis are conditioned on SST.}
\label{f1}
\end{figure}

% Figure 9:
\begin{figure}
\noindent\includegraphics[width=38pc,angle=0]{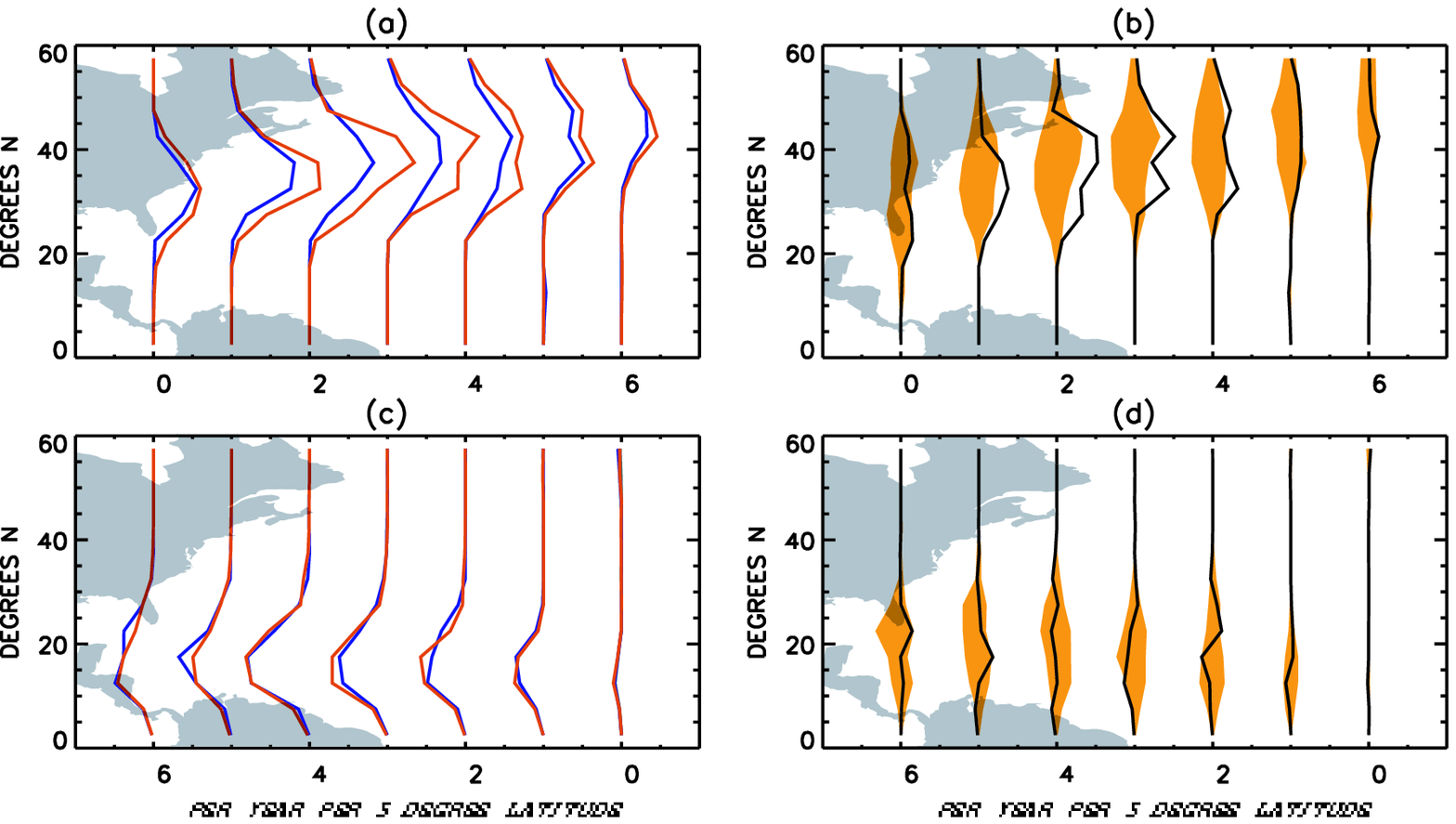}\\
\caption{As in Fig. 6, but here the hot-cold conditioning is performed for the TC propagation and lysis only.}
\label{f1}
\end{figure}

% Figure 10:
\begin{figure}
\noindent\includegraphics[scale=1]{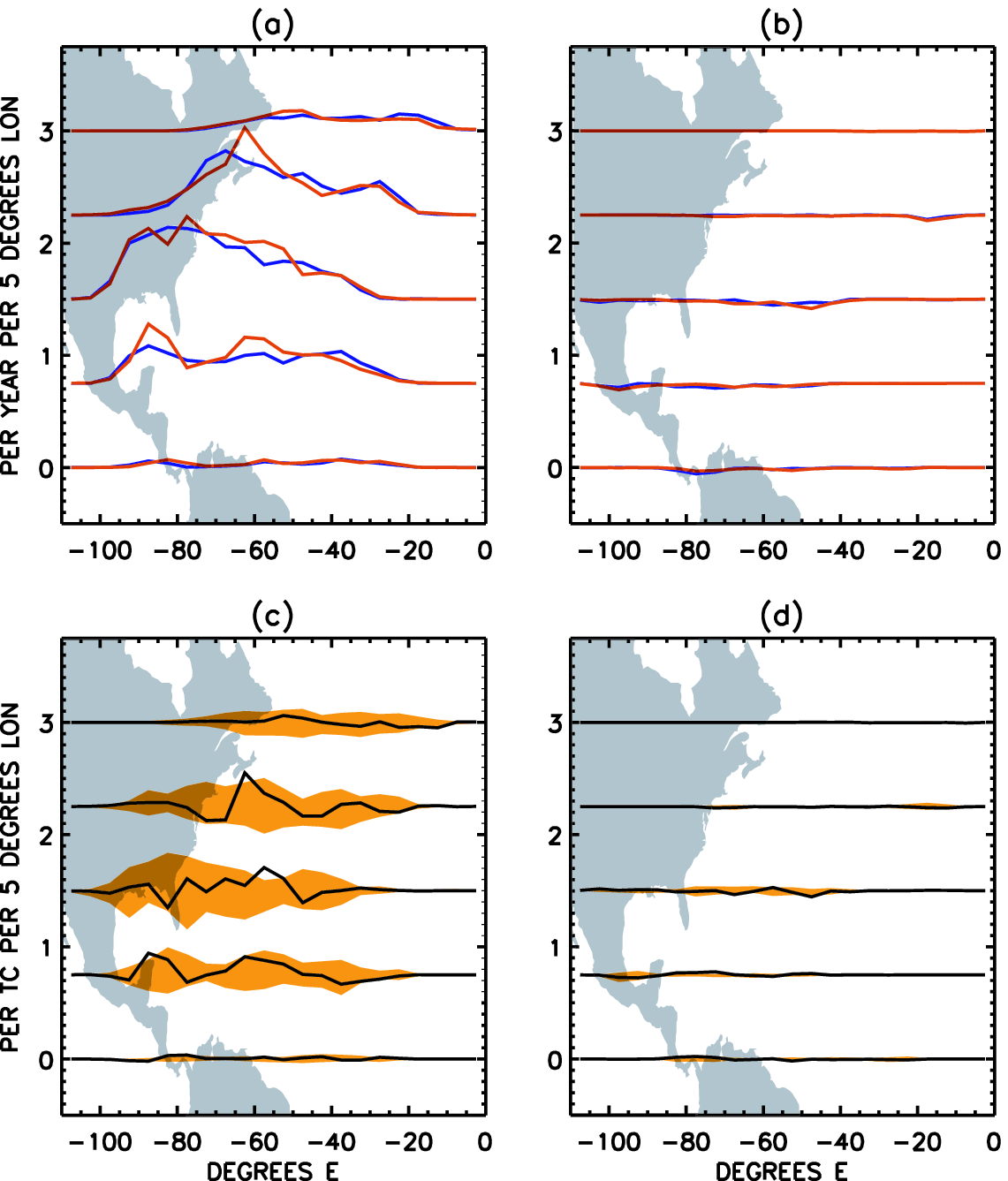}\\
\caption{As in Fig. 7, but here the hot-cold conditioning is performed for the TC propagation and lysis only.}
\label{f1}
\end{figure}

% Figure 11:
\begin{figure}
\noindent\includegraphics[scale=0.8]{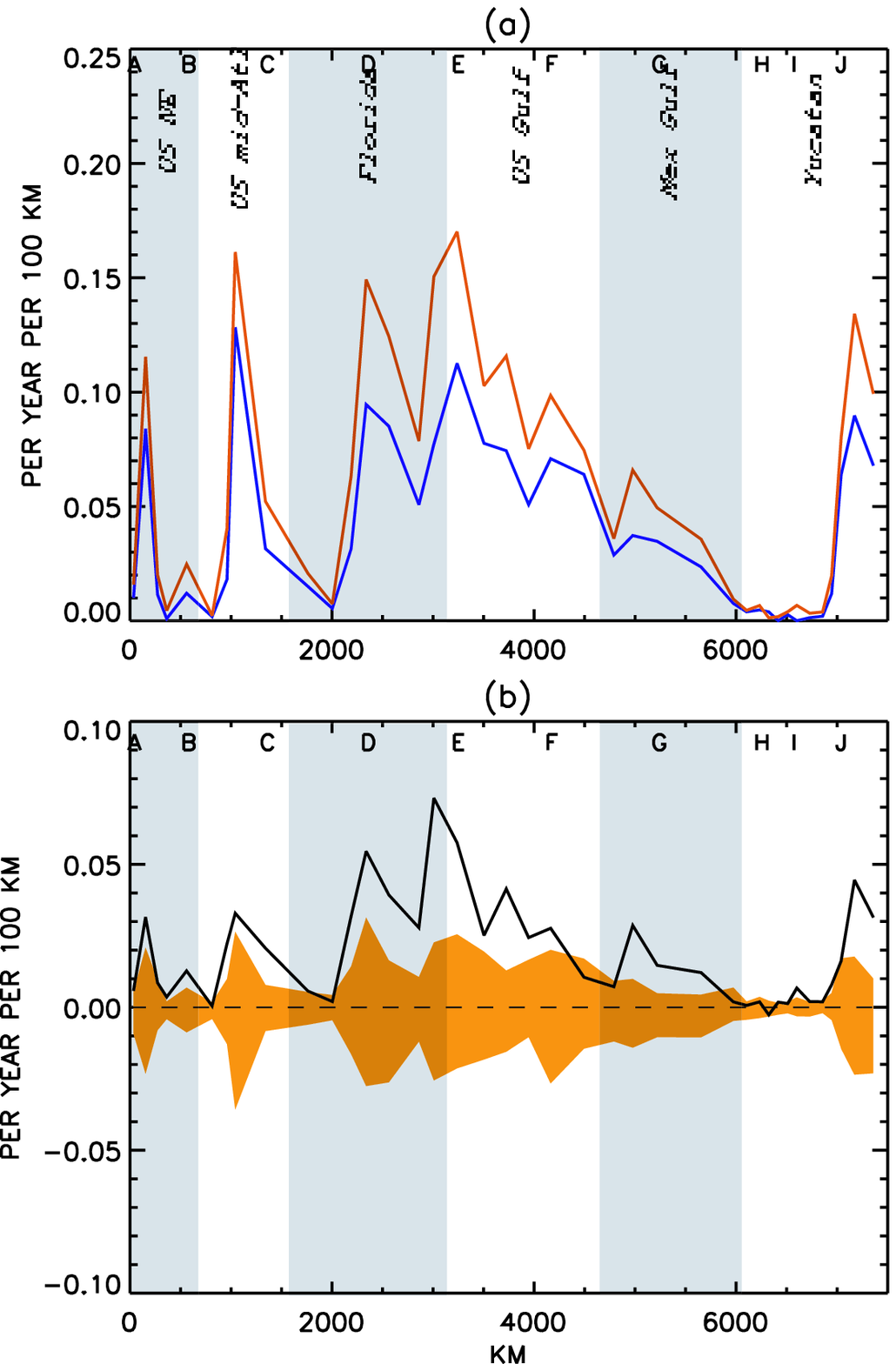}\\
\caption{(a) Landfall rates per year per 100 km of segmented coastline from Maine to Yucatan for the track model conditioned hot years (red) and cold years (blue).  (b) The hot-cold difference in landfall rate (solid) and the random-difference mean $\pm$ rms deviation difference (orange fill).  Here, only the annual TC number is conditioned on SST.  The other track model components (genesis site, propagation, and lysis) are unconditional.  Mileposts (A--J) and regions are labeled for ease of reference to Fig. 3.  Vertical shaded regions correspond to the regions defined in Fig. 3.}
\label{f1}
\end{figure}

% Figure 12:
\begin{figure}
\noindent\includegraphics[scale=0.8]{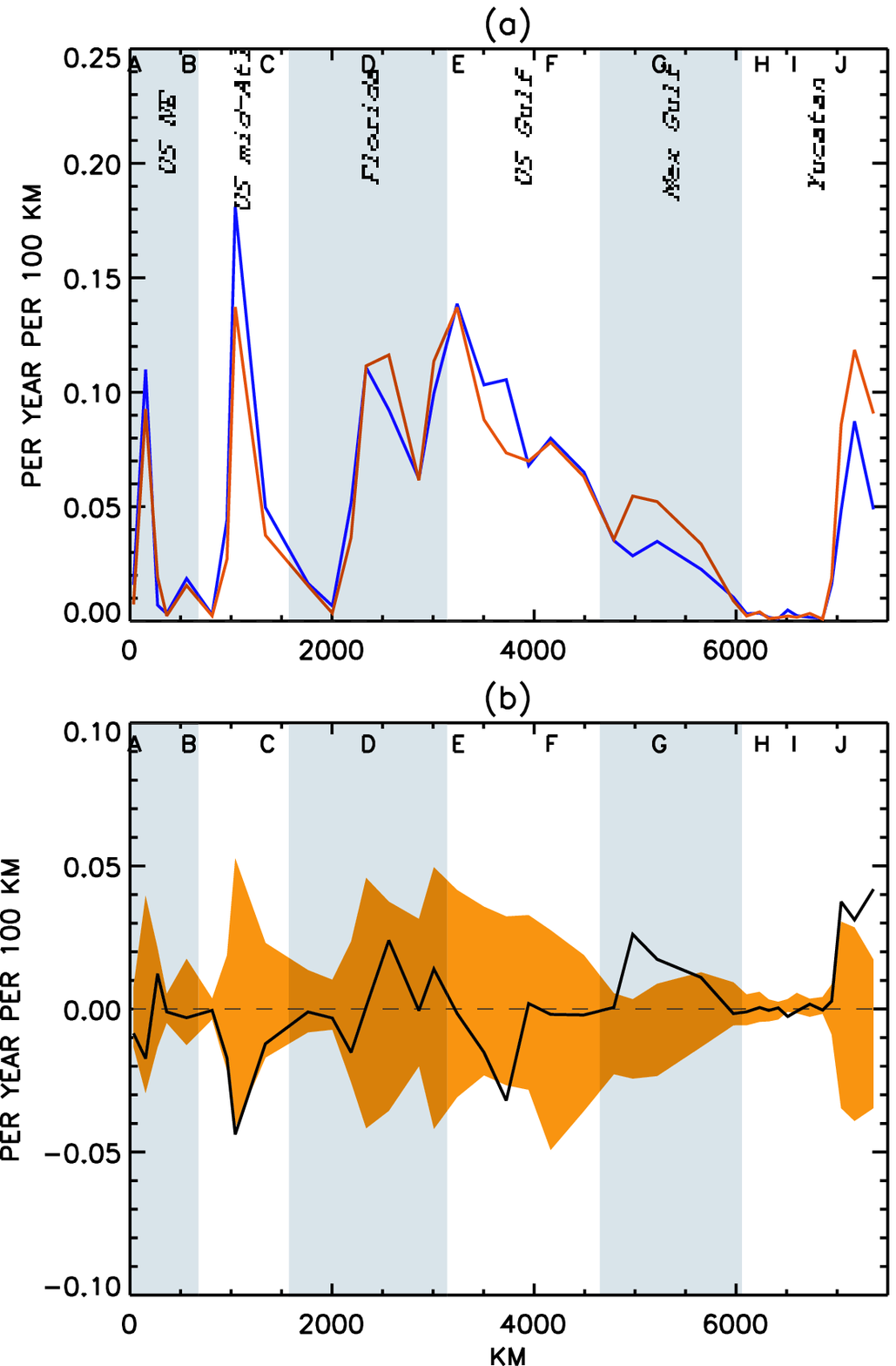}\\
\caption{As in Fig. 10, but here genesis site is conditioned on SST, and the other components are unconditional.}
\label{f1}
\end{figure}

% Figure 13:
\begin{figure}
\noindent\includegraphics[scale=0.8]{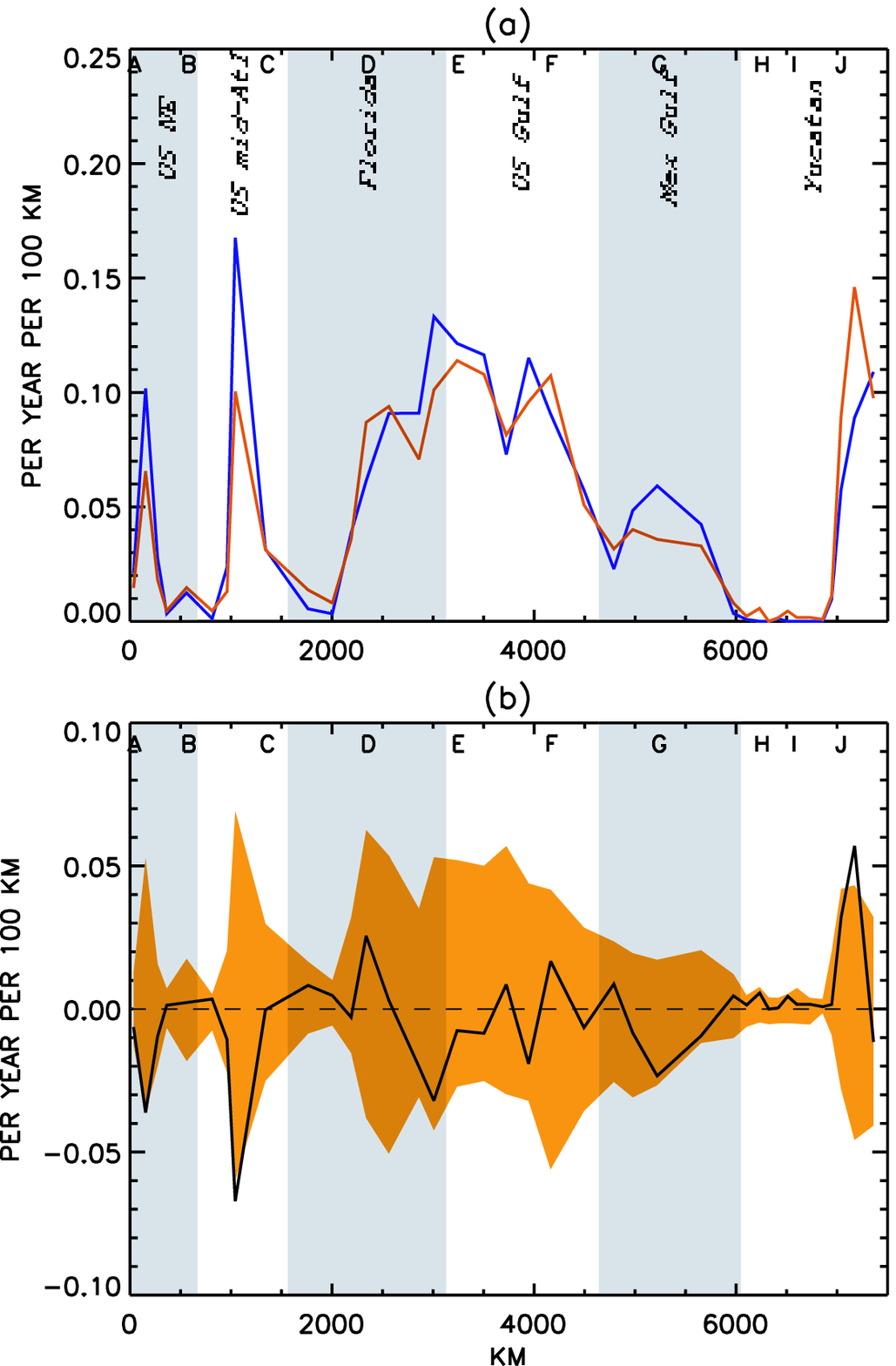}\\
\caption{As in Fig. 10, but here propagation and lysis are conditioned on SST, and the other components are unconditional.}
\label{f1}
\end{figure}

% Figure 14:
\begin{figure}
\noindent\includegraphics[scale=0.8]{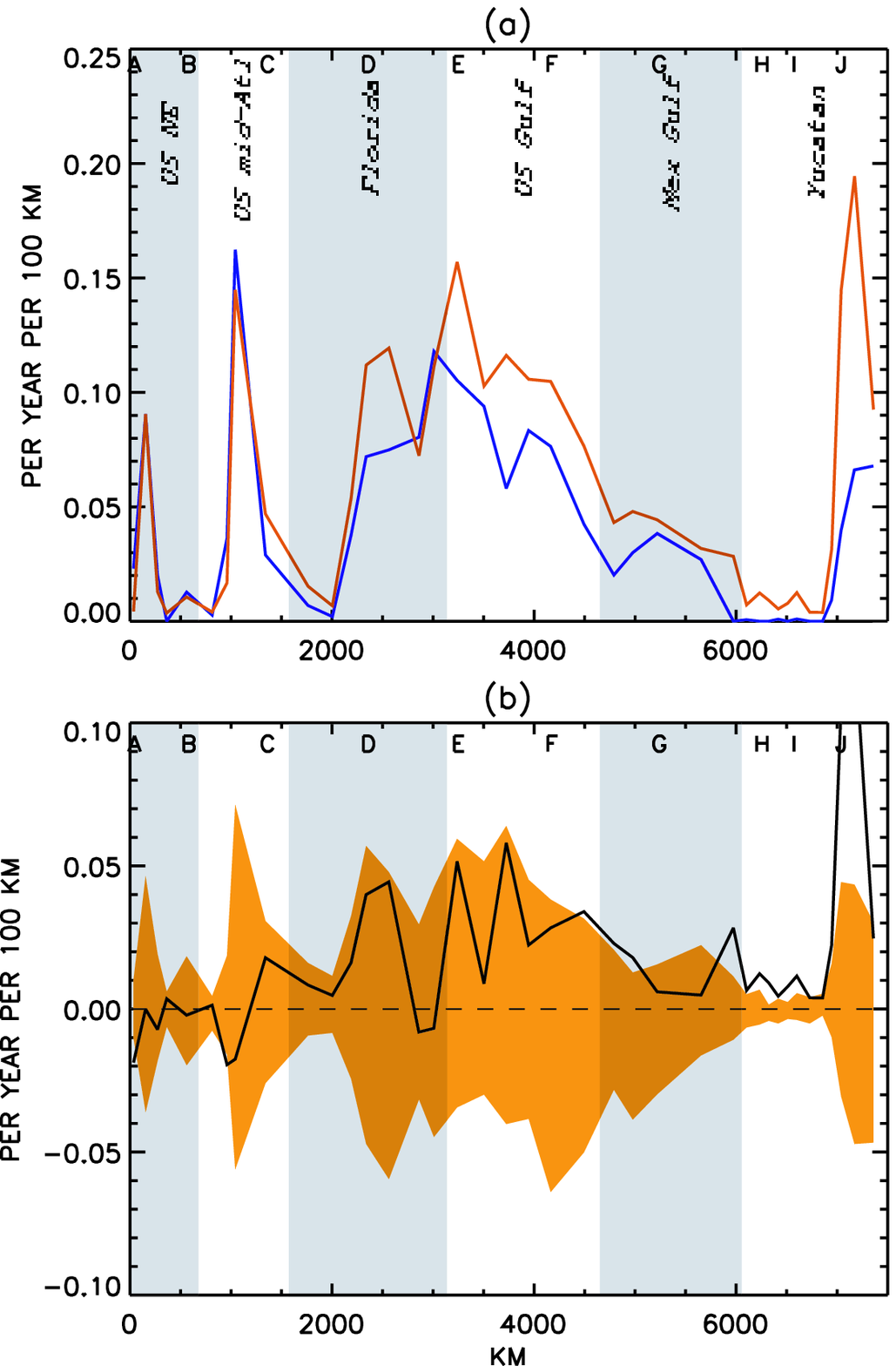}\\
\caption{As in Fig. 10, but here all components of the track model (TC annual number, genesis site, propagation, and lysis) are conditioned on SST.}
\label{f1}
\end{figure}

% Figure 15:
\begin{figure}
\noindent\includegraphics[scale=0.8]{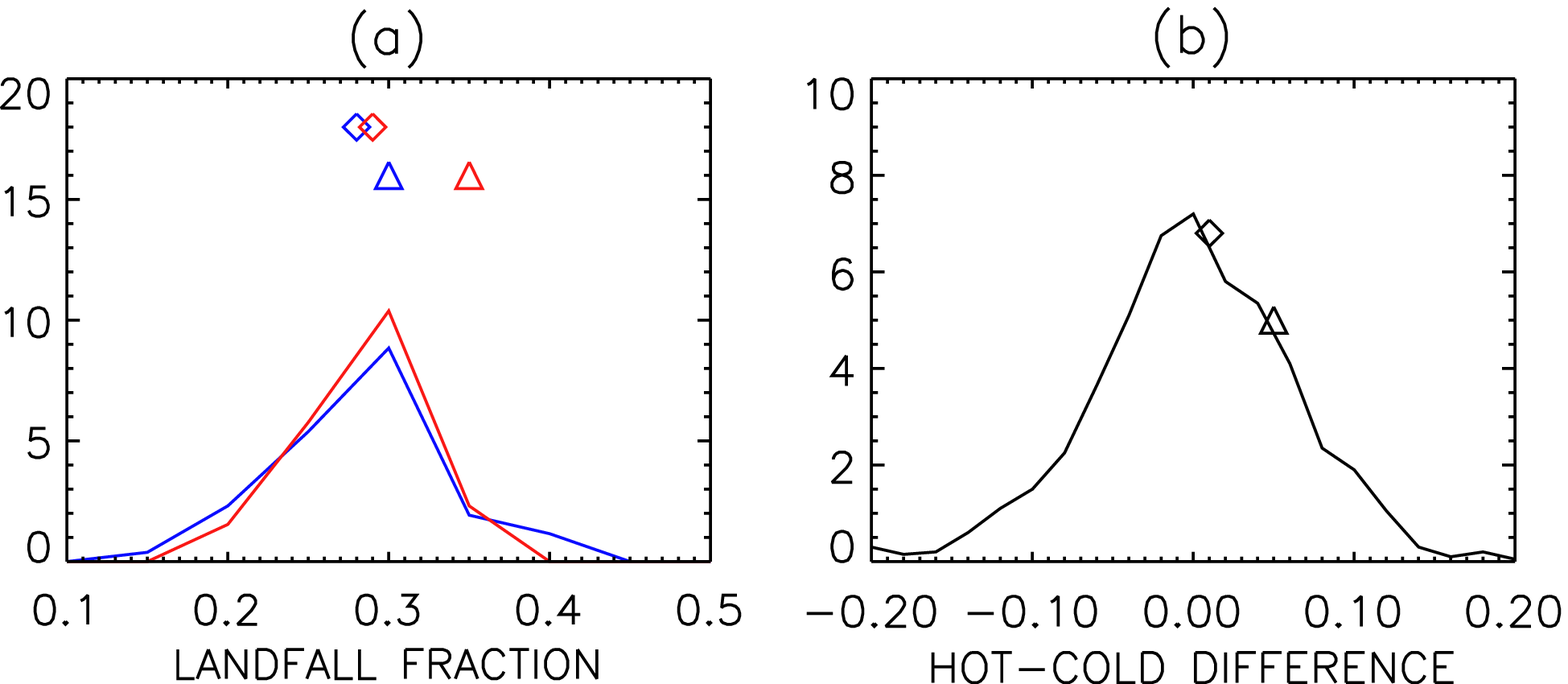}\\
\caption{(a) PDF of landfall fractions in 19-year periods computed over 52 such periods in the 1000-year cold simulation (blue) and 1000-year hot simulation (red).  Diamond symbols indicate the hot (red) and cold (blue) landfall fractions over the entire 1000 simulation.  Triangle symbols indicate the historical landfall fraction in the 19 hot years (red) and 19 cold years (blue).  (b) PDF of hot-cold landfall fraction differences obtained by differencing random samples of hot and cold simulated 19-year landfall fractions.  The diamond indicates the simulated hot-cold landfall-fraction difference and the triangle the historical hot-cold landfall fraction difference.  The analysis indicates that the historical hot-cold landfall-fraction difference is not significant.}
\label{f1}
\end{figure}

% Figure 16:
\begin{figure}
\noindent\includegraphics[scale=0.8]{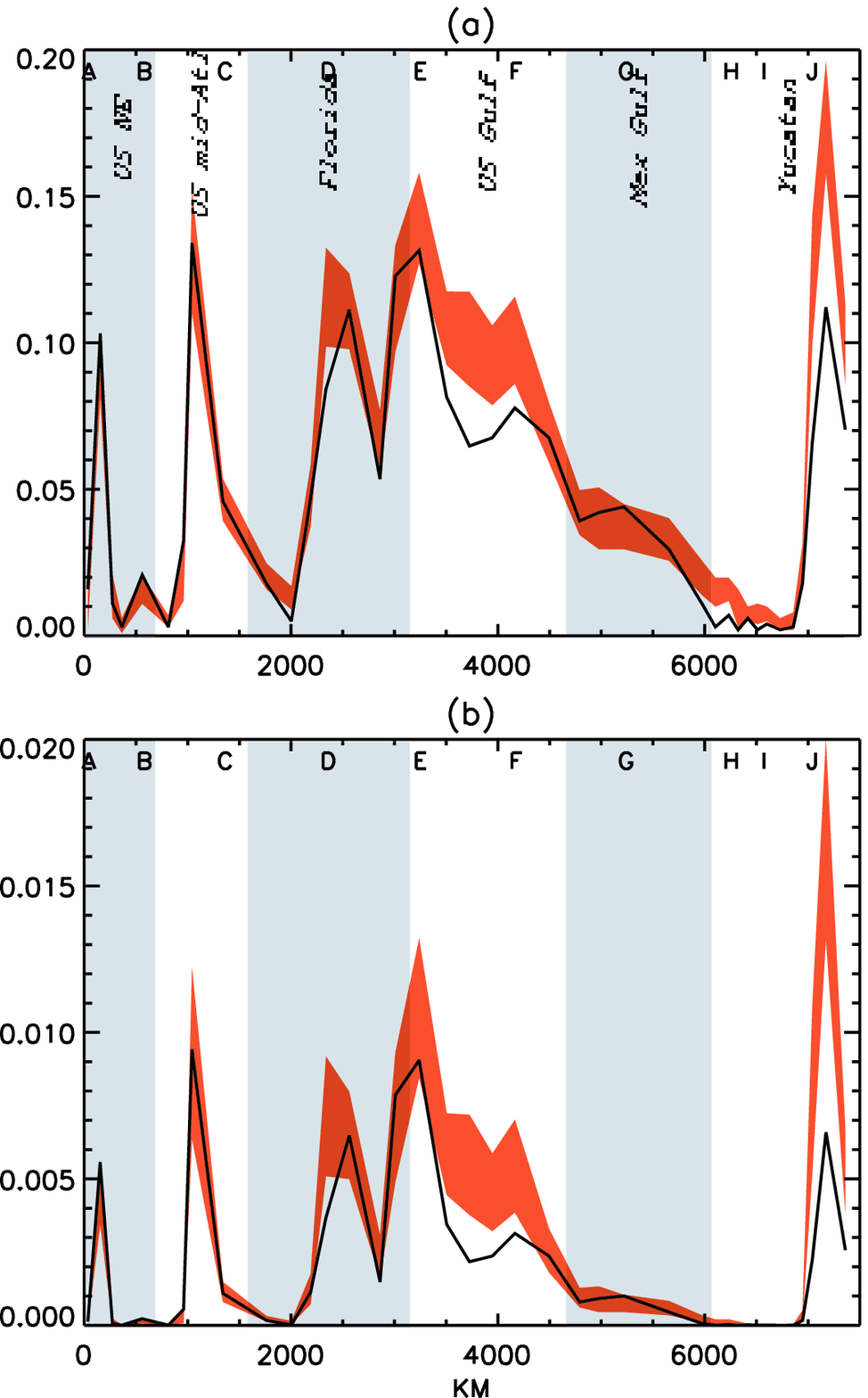}\\
\caption{Probability in 100 km sections of segmented coast of (a) at least one TC landfall and (b) at least two TC landfalls.  The probabilities for hot years are shown in red and the probabilities for all years in black.  The range in the hot year data represents the 90\% confidence limits.  These limits are determined from the 19 values resulting from successively dropping out one of the 19 hot years and computing the landfall rates and probabilities using the remaining 18 years.  Note that from (a) to (b) the y-axis is rescaled by a factor of 10.}
\label{f1}
\end{figure}

\clearpage

\begin{table}[h]
\caption{Landfall rates (counts per year) on 6 regions and the full coast for hot years ($R_H$) and cold years ($R_C$) and the $Z$ score of their difference, given by $R_H-R_C$ divided by the rms deviation of random differences.}
\begin{center}
\begin{tabular}{|c|c|c|c|c|c|c|c|}
\hline
 & US NE & US mid-Atl & Florida & US Gulf & Mex Gulf & Yucatan & Full  \\
 \hline
 $R_H$   &  0.20  &  0.39  &  0.95 &  1.84   &  0.52  &  0.75 & 4.65 \\
 $R_C$    &  0.22  & 0.33   &   0.71      &   1.22       &     0.36      &   0.31  & 3.12        \\
$Z$    &    -0.2    &    0.3  &   0.9         &     1.1        &    0.9       &  2.4  & 2.1\\
\hline
\end{tabular}
\end{center}
\end{table}

\begin{table}[h]
\caption{Probabilities of at least 1 ($P_1$), at least 2 ($P_2$), and at least 3 ($P_3$) landfalls on 6 coastline regions and the full coast for the hot years and for all years, as labeled.  Ranges on hot-year probabilities are 90\% confidence intervals derived from drop-one-out track modeling of the 19 hot years.}
\begin{center}
\begin{tabular}{|c|c|c|c|c|c|c|}
\hline
 & US NE & US mid-Atl & Florida & US Gulf & Mex Gulf & Yucatan \\
 \hline
 $P_1$(hot) &  0.17--0.21 & 0.29--0.36 & 0.59--0.65 & 0.80--0.84 & 0.34--0.42 & 0.49--0.56    \\
 $P_1$(all)    &   0.23          &   0.33        &     0.59       &  0.76         &   0.37         &  0.37       \\
 \hline
$P_2$(hot)   &  0.016--0.023 & 0.05--0.07 &  0.22--0.29  &  0.48--0.57  &  0.07--0.10 & 0.15--0.20  \\
$P_2$(all)    &  0.029             &  0.06          &  0.22           &  0.42            &  0.08          & 0.08   \\
\hline
$P_3$(hot)  & 0.001--0.002 & 0.005--0.010 & 0.06--0.09 &  0.22--0.30 & 0.009--0.018 & 0.03--0.05  \\
$P_3$(all)   & 0.002             &  0.008            &  0.06         & 0.18           & 0.012          &  0.01  \\
\hline
\end{tabular}
\end{center}
\end{table}

\clearpage

\end{document}